# Uncovering the complex mechanisms behind nanomaterials-based plasmon-driven photocatalysis through the utilization of Surface-Enhanced Raman Spectroscopies

Mahadi Hasan* and Mahamadu Tiah Mahama

Department of Chemistry, University of South Dakota, Vermillion, SD-57069, United States

**ABSTRACT:** Plasmonic materials have got wide attention as a potential candidate for light-driven catalysis of chemical conversions by harnessing solar energy to reduce the environmental issues generated by fossil fuels-based bulk chemical industries. Toward diluting this crisis, many reported the utilization of plasmonic nanostructured materials in driving industrially important reactions at normal temperature and pressure. However, these chemical conversions often suffer from low yield difficulties, and poor efficiency problems. Also, the mechanism of energy transfer from plasmon to desired molecules is still not properly understood which retards the efforts of scaling them up. Recent endeavors of using SERS to explore the complex mechanisms associated with plasmonic photocatalysis have shown great promises because of the higher sensitivity of the technique. In this article, our aim is to analyze the role of ultrafast SERS in revealing charge transfer as the primary mechanism of photocatalytic reaction of 4-nitrothiophenol to azobenzene-4,4´-dithiol and how nanoparticles other than noble metals can also drive this dimerization reaction with similar potential of noble metals. We also intend to investigate why plasmonic heating occurring at metal surface is inadequate to run photocatalytic conversions. A series of analysis has been done in these works using SERS to gather data and understand the mechanisms involved in the processes by pumping plasmonic resonance simultaneously and investigating signals from molecules at various time delays. These results open the door for researchers to consider SERS as a promising technique to uncover the physical and chemical processes involved in plasmon-driven photocatalysis and to advance toward environment friendly and cost-efficient photochemical conversions in future.

**INTRODUCTION**: Industrial chemical processes consuming 35% energy in the United States results in the emission of around 280 million tons of carbon-di-oxide every year which anticipated to be doubled by next 30 years.[4] The incorporation of less energy consuming chemical reactions using renewable materials is therefore important to address the impacts of carbon emission in the nature. Toward achieving this goal, photocatalytic reactions with plasmonic nanomaterials to utilize solar energy for driving chemical reactions at standard conditions could play a vital role. Some instances of these reactions are intramolecular alkyl transfer,[6] cycloaddition of azide-alkyne,[7] dimerization of p-nitrobenzenethiol,[5] and oxidation-reduction reactions.[8] Although these reactions produce final products in low scale compared to bulk industrial production, large-scale production for industrial chemical synthesis is very possible by tuning different properties of plasmons. Nevertheless, the lack of comprehensive study of mechanisms involved in plasmon-analyte interactions has impaired the perfect utilization of these substances so far.

The conduction band electrons in plasmonic metal nanostructures oscillate collectively due to localized surface plasmon resonances in the response of incident light to enhance electromagnetic field around nanoparticles surface which helps in efficient conversion of solar energy to chemical energy.[9,25] The interaction between plasmonic light and nanomaterials create single molecule like nanoscopic areas called hotspot. These hotspots become very active to start chemical reactions using electron-hole, heat and light energy after LSPR degradation and energy splitting. The physical processes which may play a vital role to drive reactions are heating, charge migration, electromagnetic field enhancement etc. It is significant to explore the key mechanisms for driving these photochemical reactions with high production rate and selectivity.

Charge migration of plasmonic materials from energized hole-electron to adjacent molecules involved in redox reactions has great importance in photocatalytic processes. In the case of indirect mechanism, hot carriers are transferred from the surface of plasmonic nanomaterials to vacant orbitals of nearby molecules.[8] However, the gradual cooling of hot electrons through thermal scattering makes this migration method inefficient. On the other hand, direct charge migration involves direct excitation of metal-analyte through hot carriers in where the orbitals of both metal and analyte molecules are hybridized and the plasmon resonance energy matches with the energy difference of the nanomaterial-analyte pair.[10,12]

Heating of plasmonic materials through LSPR decay also has a significant role in understanding mechanisms of the plasmon-mediated photocatalysis processes. Plasmonic heating supplies sufficient heat energy to adjacent adsorbates which helps them to exceed the threshold energy and thereby run the chemical reaction smoothly.[1] However, heating gives subsidiary role in these reactions because the nanomaterials generating heat are expensive and therefore finding the rational design of selectivity for heat driven reaction is difficult.

Non-noble nanostructured metal plasmonic substances has recently got wide attention in photocatalytic reactions with a view to attenuating the expense of noble metals as well as harnessing unique characteristics of plasmonic materials.[3] However, the mechanism of driving photocatalysis by these non-noble metals remains somewhat unclear compared to their noble metal counterparts.

It is hard to realize the mechanism or combination of mechanisms like charge migration, heat transfer working behind photocatalytic reactions to make them highly efficient. On top of that, higher sensitivity of the methods employed is necessary since light induced molecular rearrangements occur in picosecond-femtosecond time periods. Toward obtaining this goal, SERS could be the best fit to gather details about the mechanisms involved in photocatalysis by determining molecular vibrations. SERS can introduce remarkable detection sensitivity by the enhancement of signals which can be up to $10^{12}$ times in a properly optimized analyte.[14] SERS has high sensitivity to almost all chemically reactive species because most of the signals arise from the analytes adsorbed in the hotspots. In this review, the role of ultrafast SERS will be discussed in exploring the processes associated with plasmon-driven photocatalytic conversion reactions.

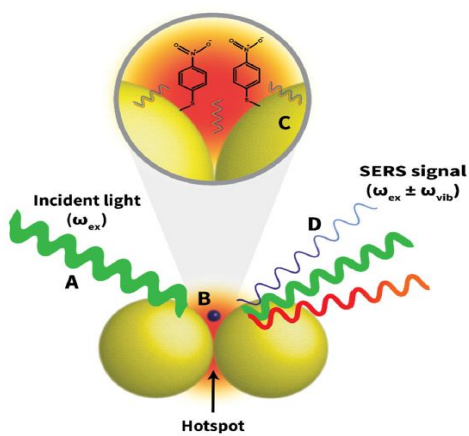

*Figure 1. Major events in a SERS analysis A) Excitation of the LSPR. B) Electromagnetic area enhancement in hotspots. C) Plasmon-photon interaction. D) SERS signals*[1,25]

**METHODS:** Ultrafast Surface-Enhanced Raman Spectroscopy employed to identify the mutual interactions between molecules and plasmon on a picosecond course of time yields enough insights on molecular vibrations due to LSPR for real-time analysis.[1,2,3] In this work, plasmon excitation was done by using a femtosecond pumping pulse and a picosecond probing pulse was employed to observe the transitions in molecular configurations at different time intervals. The instrument response was determined by Ker effect using optical behavior over varying picosecond and femtosecond time scales. Ultrafast spectroscopy using lasers of 2-24 MHz rate of repetition having 10 MW/cm$^2$ flux was employed for the mitigation of sample breakage of metal nanomaterials due to bigger flux of lasers.[15-17] Some physical measures like stirring and rotating substrates at high rpm were used to achieve stable samples avoiding the degradation for colloidal and solid samples respectively. In the active mode of the spinning stage, signal enhancement is seen for Au nanofilm which undergoes surface reconstructions over the nanosphere whereas substrate ablation occurs by the effect of intense laser pulses during inactive mode and molecules start to be released from the surface. These revolutionary results showed the way to move forward for the study of mechanisms involved in plasmonic heat generation, tracking the hot electrons in extremely short time periods, optical factors controlling the chemical activeness of plasmonic materials.

**RESULTS AND DISCUSSIONS**

**MODE OF CHARGE MIGRATION**

A myriad of plasmon driven photocatalytic reactions are known to follow the mechanism of the transfer of hot electrons from nanostructured noble metals to adsorbate molecules. Some important elements such as electron delocalization, creation of phonon (quantized vibrations in metal lattice), orientation of adsorbate, hybridization between noble metal and adsorbate governs these charge migration mechanisms (direct or indirect). [20-22] In this work, the Frontiera group used ultrafast SERS to perfectly explore these mechanisms behind the dimerization reaction of *p*-NBT on Ag metal surface since LSPR decay is very quick and hot carriers have very brief lifespan.[1]

Since hot electrons play major role in driving the dimerization reaction of *p*-NBT, ultrafast SERS was set to action to study the dynamics of these hot electrons in the system.[23] This reaction occurs at standard light through the coupling of reporter molecule *p*-NBT on the surface of aggregated silver NPs. As shown in Figure 2, the SERS spectra indicate the near-field interaction between photons from Stokes shift and nanoscopic photoluminescence of hot carriers created from LSPR decay,[24] this phenomenon is known as Fano resonance. They found the lifetime of SERS signals ~3.9 picoseconds which matches with standard photoluminescence lifetime of electron. These analyses indicate that the peak amplitude for *p*-NBT is higher than that of DMAB in the ground state while the trend is totally opposite for the transient state. This increase in the product peak amplitude is the result of deexcitation of conduction band electrons of metals to valence band, thereby creation of hot electrons. The pathways for electron transfer of plasmon-adsorbate interaction can be indirect or direct. The results from ultrafast SERS spectra give the idea of indirect electron transfer mechanism in this reaction because photoluminescence is observed on the surface of metal where hot electrons are deposited. These observations pave more ways to study the use of ultrafast SERS technique in exploring the impacts of morphology modification towards rational design of light induced photochemical reactions.

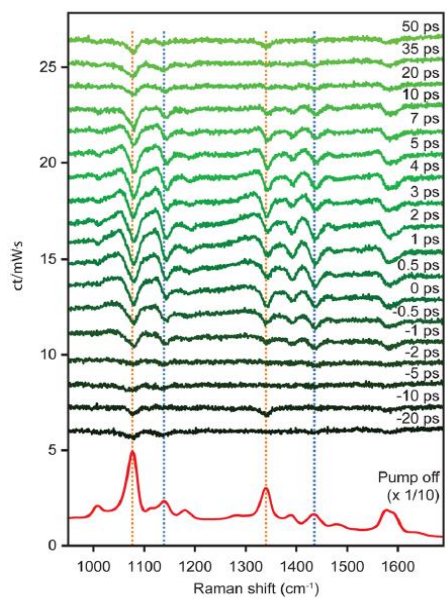

*Figure 2. Different spectra obtained from ultrafast SER analysis of p-NBT (blue dotted peaks) and DMAB (orange dotted peaks).[1]*

**SUBSIDIARY ROLE OF HEATING**

Majority of the plasmon mediated reactions are endothermic.[30] The temperature rise in hot carriers after plasmonic degradation can go up to several thousand kelvins for electrons and hundreds of kelvins for phonons in nanostructured metal plasmons.[18] However, it is matter of great debate whether heating is major mechanism for plasmon-driven photocatalysis or not.

In this study, the Frontiera group used ultrafast SERS along with Raman thermometry to determine the differential alteration in the effective temperature of *p*-nitrobenzenethiol and 4,4´-vinylenedipyridine on the roughened surface of Au NPs.[2] They measure the effective temperature from the change in energy of vibration using Boltzman distribution analysis by calculating Raman spectra (Stokes and anti-Stokes) over ultrashort time. This work shows that the transfer of energy from plasmon to molecule occurs in picosecond time and the change found in effective temperature is under 100K. As shown in Figure 3A, this heat spreads out from the adsorbate molecules after five picoseconds of excitation at 500 W/cm$^2$ flux for *p*-nitrobenzenethiol. Figure 3B shows the qualitative consistency of this result for every mode of vibration. They also employed a different adsorbate, 4,4´-vinylenedipyridine and found the same observations (illustrated in figure 3C). These results indicate that the temperature of adsorbate molecules is unable to travel past the energy barriers of the plasmonic reactions and the energy remains in action for a very short period. So, it is now obvious that heating from LSPR decay is not the primary driving force in light induced photocatalysis. However, this work opens the door for further studies about selectivity of nonthermal plasmon mediated mechanisms using ultrafast SERS.

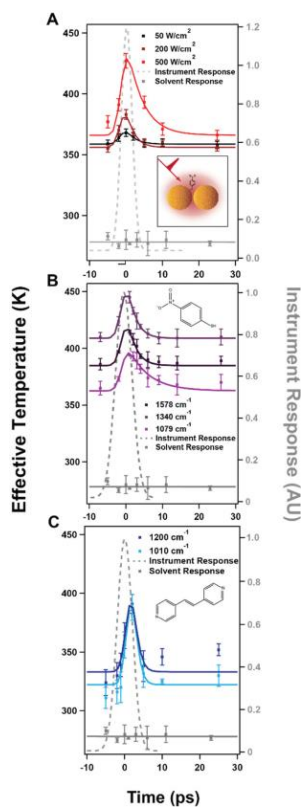

*Figure 3. Energy allocation patterns at 1035 nm laser excitation- (A) of 1079 cm$^{-1}$ vibrational mode for p-NBT when EF measured at various flux of excitation, (B) of three different modes for the same molecule, and (C) of two different modes for a different molecule 4,4'-vinylenedipyridine adsorbed on gold NPs aggregate.[2]*

**DOPED NANOPARTICLE PLASMONS**

The facile sol-gel preparation methods,[26,27] and unique characteristics of plasmonic nanomaterials based on noble metals are widely recognized and their applicability is growing fast with technological advances. However, noble metal nanomaterials are rare, costly, and show limited tendency to other elements which hinders the versatile use of them. Therefore, recent focus has been shifted towards utilizing readily available and cost-effective non-noble metal nanomaterials to expand the exceptional application of noble metals.[11] Like noble metal-based nanomaterials, the chemical and physical mechanisms of photocatalysis for non-noble metals are still under investigation.

In this article, they reported preparation of DMAB from *p*-nitrobenzenethiol for the first time, photocatalyzed by doped copper selenide nanoparticles and the yield of the reaction is almost like Au catalyzed reactions.[3] They used SERS to measure the light induced enhancement of the electromagnetic field to compare plasmonic behavior of copper selenide nanoparticles with typical noble metal plasmons. They first determined a tentative estimate of SERS enhancement factor (EF) by using benzenethiol as the reporter molecule. As shown in Figure 4A, the intensity of Raman peaks increases as the concentration of the reporter molecules goes high. The EF of copper selenide nanomaterials was measured by the equation, EF= $I_{SERS} N_{NRS}/N_{SERS} I_{NRS}$; using the data from the linear regression of the plot of $I_{SERS}/N_{SERS}$ vs $N_{EXCESS}$ (Figure 4B). Here, $I_{NRS}$ and $I_{SERS}$ denotes respectively the peak intensities of benzenethiol for traditional Raman analysis and SERS at 1002 cm$^{-1}$ while the number of reporter molecule taken in the bulk solution and surface of copper selenide are indicated by $N_{NRS}$ and $N_{SERS}$ respectively. For Cu$_{2-x}$Se nanoparticles, the

SERS enhancement factor was found around $10^4$ while the standard value lies in the range of $10^3$-$10^7$ for conventional noble metals.[13]

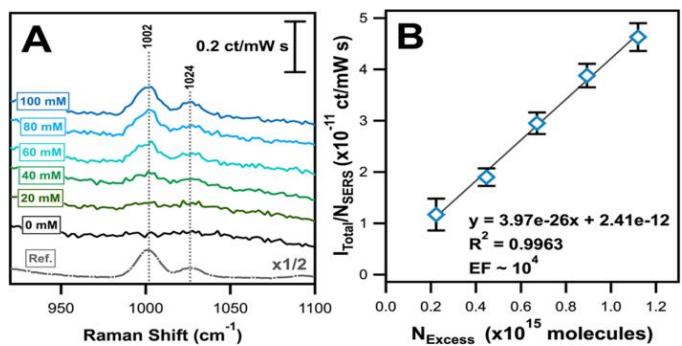

Figure 4. (A) Intensity of Raman peaks as concentration of reporter molecule adsorbed on copper selenide increases (B) ISERS/NSERS vs NEXCESS graph.[3]

After getting the SERS EF measurement, they did investigate on how non-noble metal LSPR can drive the dimerization reaction of 4-nitrothiophenol to azobenzene-4,4´-dithiol. As shown in Figure #, the Raman peak for the product of the reaction has been moved by 10 cm$^{-1}$ to 15 cm$^{-1}$ from that of noble metal mediated reactions which clearly indicates the difference in chemical ambience and the binding patterns on copper chalcogenide nanoparticles surface. But the mechanism of LSPR decay develops through the hot holes instead of hot electrons in this case. So, though non-noble metal plasmon-mediated photocatalysis opens the doors of huge possibilities compared to noble metal counterparts, further experimental or computational studies are required to completely understand the light induced reaction mechanisms and dynamics of the LSPR decay.

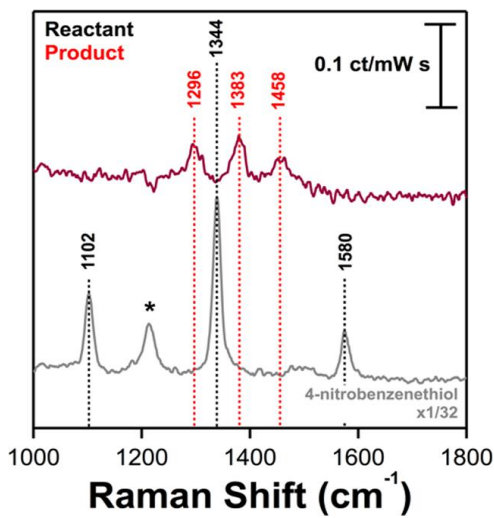

Figure 5. Surface-Enhanced Raman spectra for the evolution of DMAB peaks (red) from p-NBT (black) adsorbed in copper selenide NPs within 5 minutes of irradiation.[3]

**CONCLUSION AND OUTLOOK:** Taking everything into account, though there is need for supplementary refinement of the study of photocatalysis through the plasmonic nanomaterials to make them commercially achievable on a larger scale, the discussed SERS strategies have potential insight to obtain this goal. The usefulness of steady state as well as ultrafast SERS analytical methods in unravelling the heating pattern, electronic shifts, and vibrational details for photocatalytic reactions by plasmonic substances is successfully described in the reviewed articles here. So, it is crucial to focus on the development of various potential SERS techniques for the study of the mechanisms involved in the change of resonance energy to understand the features of any plasmon mediated photocatalytic reaction properly. A close collaboration between materials chemists and theoretical researchers will facilitate tuning the features of plasmonic nanomaterials using SERS to make them capable of harnessing solar energy in cost-effective ways without doing any harm to the environment.

## AUTHOR INFORMATION

### Corresponding Author

Mahadi Hasan – Department of Chemistry, University of South Dakota, Vermillion, SD 57069; Phone: 605-202-0843; Email: Mahadi.Hasan01@coyotes.usd.edu

## ACKNOWLEDGMENTS

The author thanks Levi Spencer for reviewing the article.